\documentstyle[prl,twocolumn,aps]{revtex}
\begin{document}
\draft

\tolerance50000 %%%%\preprint{
%%%%\begin{minipage}[t]{1.8in}
%%%%\hfill LPQTH-94/?
%%%%\end{minipage}
%%%%}

\twocolumn[\hsize\textwidth\columnwidth\hsize\csname @twocolumnfalse\endcsname

\title{NMR imaging of the soliton lattice profile in 
the spin-Peierls compound CuGeO$_3$}
\author{M. Horvati\'{c}$^{1}$, Y. Fagot-Revurat$^{1}$, C. Berthier$^{1,2}$, 
G. Dhalenne$^{3}$, A. Revcolevschi$^{3}$  
}
\address{
$^{1}$Grenoble  High  Magnetic  Field  Laboratory,  CNRS and MPI-FKF, B.P. 166, 38042 Grenoble Cedex 09, France \\
$^{2}$Laboratoire de Spectrom\'{e}trie Physique, Universit\'{e} J. Fourier, B.P. 87, 38402 Saint-Martin d'H\`{e}res, France\\
$^{3}$Laboratoire de Chimie des Solides, Universit\'{e} de Paris-Sud, 91405, Orsay, France 
}
\date{\today}
\maketitle
\begin{abstract}
\begin{center}
\parbox{14cm}{
In the spin-Peierls compound CuGeO$_{3}$, the commensurate-incommensurate transition
concerning the modulation of atomic position and the local spin-polarization
is fully monitored at $T=0$ by the application of an external magnetic
field ($H$)  above a threshold value $H_{c}\simeq $ 13 Tesla. The solitonic profile
of the spin-polarization, as well as its absolute magnitude,  has been
precisely imaged from $^{65}Cu$ NMR lineshapes obtained for 
$h=(H-H_{c})/H_{c}$ varying from 0.0015 to 2. This offers a unique possibility
to test quantitatively the various numerical and analytical methods
developed to solve a generic Hamiltonian in 1-D physics, namely strongly
interacting fermions in presence of electron-phonon coupling at arbitrary band
filling.
}
\end{center}
\end{abstract}

\pacs{
PACS numbers: 75.10.Jm, 75.30.Fv, 75.50.Ee, 76.60.-k}
]

The spin-Peierls (SP) transition observed in quantum spin 1/2
antiferromagnetic chains \cite{Bray} is the magnetic analogue of the
well-known Peierls transition in quasi-one-dimensional (1D) metallic chains.
A 3D array of chains of spins 1/2, in which the position dependent
antiferromagnetic exchange interaction $J$ between neighboring spins
provides a coupling between spins and lattice distortion, i.e., phonons,
undergoes a second order phase transition at $T=T_{SP}$ from a high
temperature uniform, paramagnetic and gapless phase, to a low temperature
phase in which the lattice is dimerized. The corresponding collective
singlet ground state is separated by a gap from triplet excitations and the
cost in elastic energy is compensated by the gain in magnetic energy
(instead of electronic kinetic energy in the Peierls transition). Applying
the canonical Jordan-Wigner transformation, the Hamiltonian of such a chain
placed in an external magnetic field $H$ can be mapped onto that of
interacting (spinless) fermions coupled to the phonons, $H$ playing the role
of the fermionic chemical potential. Physical realizations of SP systems
thus offer the unique possibility to check the understanding of this
fundamental 1-D Hamiltonian for arbitrary band filling, and particularly
close to half-filling\ (corresponding to $H$ = 0) where commensurability
effect are expected to play an important role through the unklapp processes 
\cite{Brazov}. Until recently, physical realizations of SP system were
limited to organic compounds \cite{Bray}. The recent identification of the
first inorganic compound of this type \cite{Hase}, namely CuGeO$_{3}$, and
the synthesis of large and high-purity single crystals, allowing for example
inelastic neutron scattering, has therefore raised an important experimental
effort \cite{Boucher}. In particular, one of the key features of the SP
systems is the possibility by applying a strong enough magnetic field to
cancel the gap between the singlet ground state and the lowest triplet
excitation. This induces a transition to a new magnetic phase in which the
periodicity of the spin-polarization (i.e., the local magnetization) and of
the associated lattice deformation is incommensurate (IC) with the
underlying crystallographic lattice. This commensurate-incommensurate (C-IC)
transition takes place by the introduction of a soliton lattice. Each
soliton corresponds to a rapid change of phase between two equivalent
commensurate (dimerized) domains, and is predicted to carry a total spin $%
\frac{1}{2}$\cite{Nakano1}. Increasing the magnetic field above the
threshold value $H_{c}$, the modulation of the lattice distortion and the
local spin polarization will continuously evolve from a well-defined soliton
lattice to planewave regime.

In this letter, we present a detailed study of $^{65}$Cu NMR lineshapes
obtained in a CuGeO$_{3}$ single crystal in the field range 13-26 T which
gives a direct and detailed access to the real space profile of the spin
polarization in the IC phase of CuGeO$_{3}$ for $h=(H-H_{c})/H_{c}$ varying
from 0.0015 to 2. The spin polarization profiles have been successfully
parametrized using Jacobi elliptic functions, which are the generic
solutions of the sine-Gordon equation, allowing the determination of both
the period and the shape of the solitons in the whole range of $h$. These
results provide a stringent test for analytical and numerical solutions of
the Heisenberg SP Hamiltonian, enabling to investigate the influence of
second neighbors interactions, 3-D coupling, or phasons.

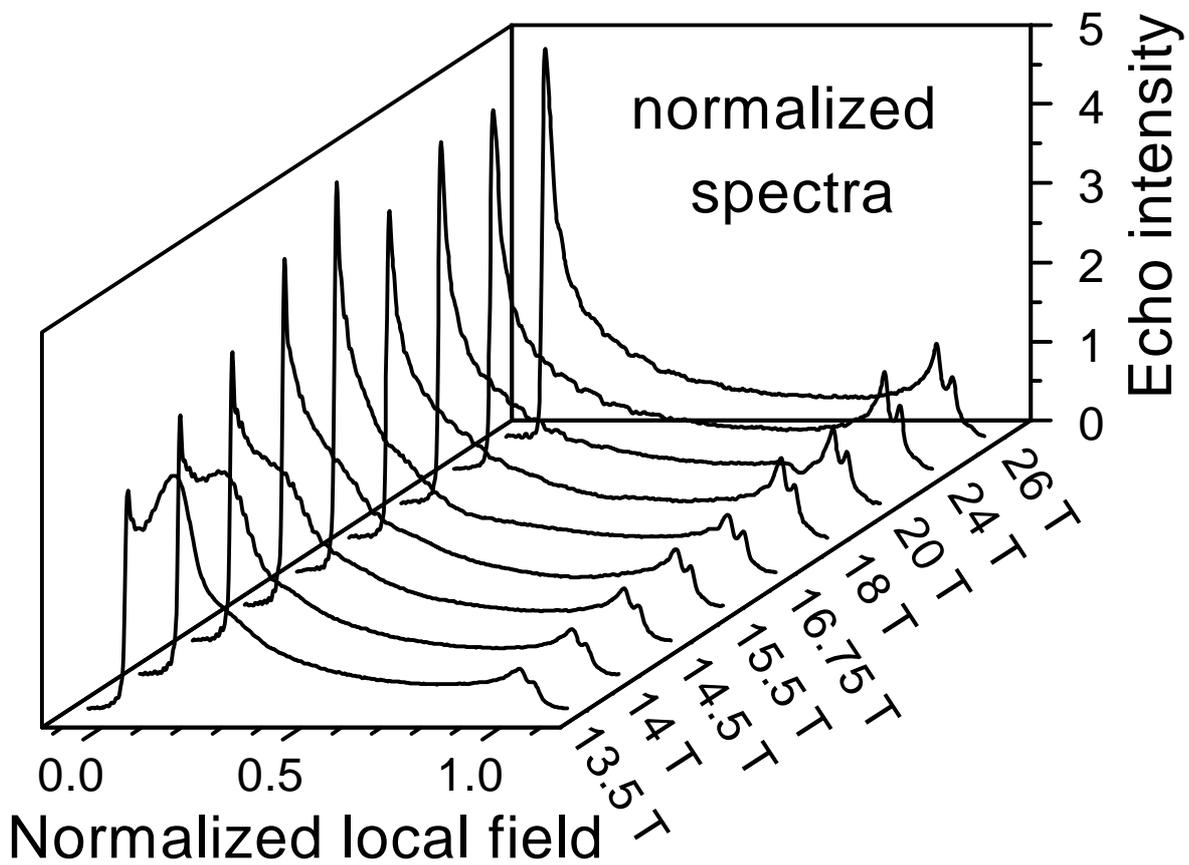
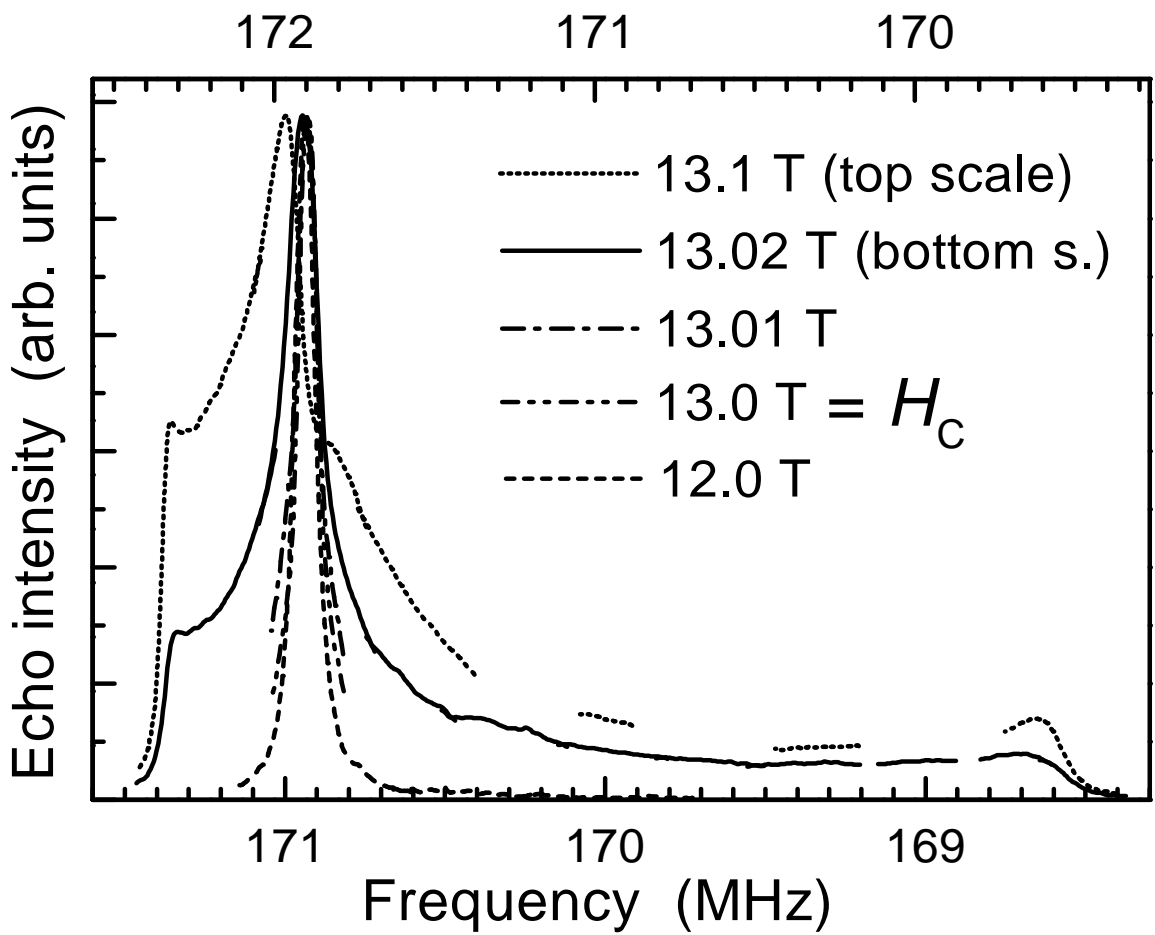
\begin{figure}[tbp]
\caption{Magnetic field dependence of $^{65}$Cu NMR lineshape in a CuGeO$%
_{3} $ single crystal at 4 K for $h=(H-H_{c})/H_{c}$ varying in the range
0.0015-2. In the bottom panel, the spectra have been recorded by sweeping
the frequency. The frequency scales of these spectra are mutually shifted in
order to compensate for the variations of magnetic field. In the top panel,
the spectra correspond to field sweeps.}
\label{Fig1}
\end{figure}

In Fig. \ref{Fig1} is shown the evolution of Cu NMR lineshapes when $H$ is
varied from $H_{c}$ to $\simeq 2H_{c}$. The data are taken on the high
frequency satellite of $^{65}$Cu isotope in order to avoid overlapping of
the lines and ensure maximal resolution. All spectra have been taken at $T=$
4 K and for $H$ parallel to the $c$-axis of the single crystal, {\it i.e.,}
along the chains direction. Reducing $T$ to 1.3 K would increase the width
of the spectra by only 5 \% while keeping the same lineshape. For values of $%
H$ close to $H_{c}$ = 13 T, the spectra have been recorded at fixed
value of $H$ and sweeping the NMR frequency (Fig. \ref{Fig1}, bottom
panel), and for higher values of $H$ at fixed frequency sweeping the
magnetic field (top panel); the latter method is technically more convenient
for wide lines, and is used whenever the modification of the line-shape
during the sweep is negligible. Splitting of the high field edge singularity
in these spectra is due to slight misalignment of the single crystal. Up to
the critical field $H_{c}$, only a narrow symmetric line is seen,
corresponding to the homogeneous zero spin polarization in the dimerized
phase. Just above, this line is strongly broadened and a wide distribution
due to solitons appears. With increasing $H$, the broadened zero spin
polarization peak is rapidly suppressed, reflecting the disappearance of the
nearly dimerized regions and the approach towards a planewave-like regime.

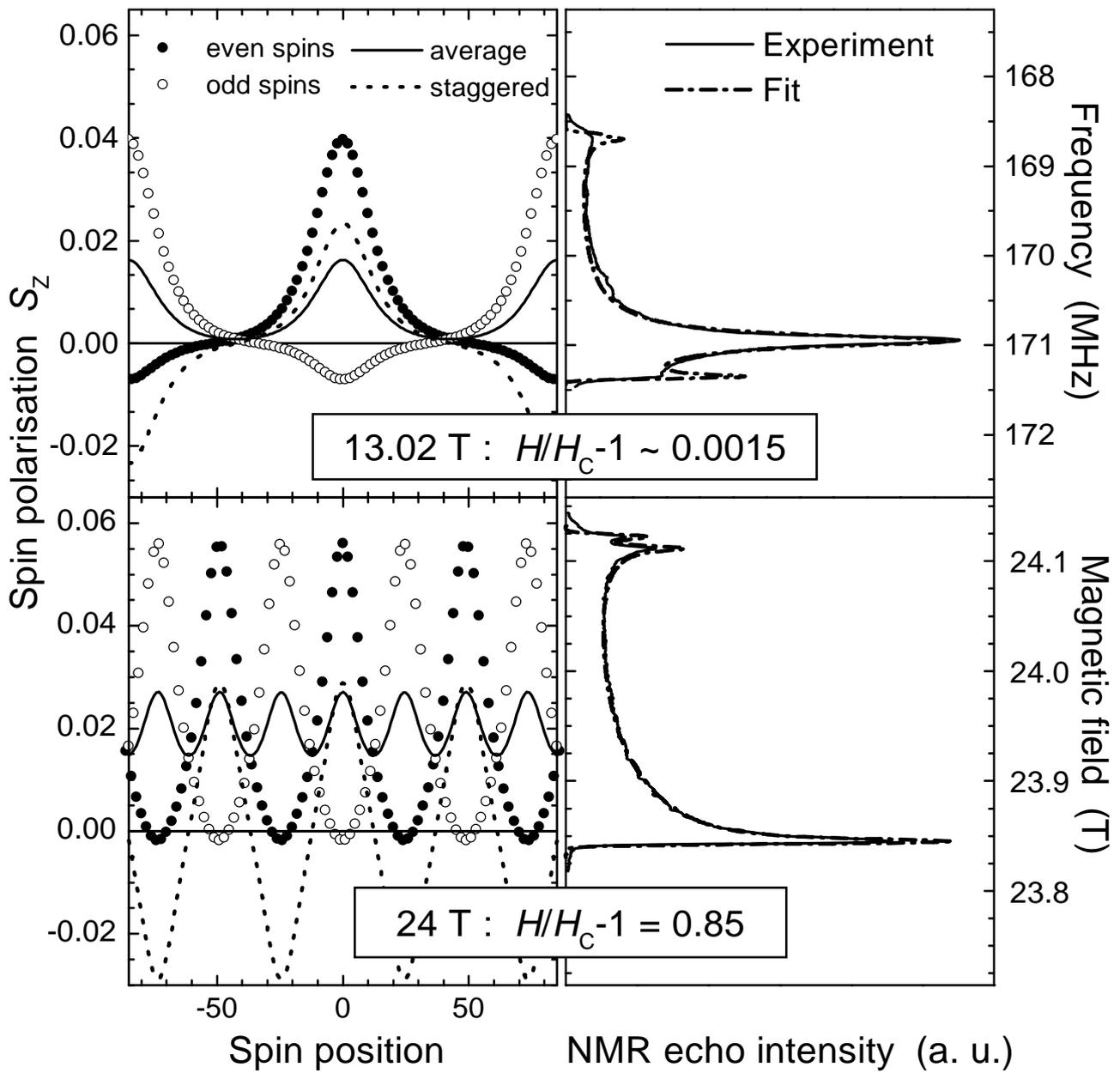
\begin{figure}[tbp]
\caption{Real space reconstitution of the spin polarization profile (left)
and the corresponding fit to Jacobi elliptic functions superposed on
experimental spectra (right), as explained in the text. Left top panel shows
one period of soliton lattice taken at field only 0.15 \% above $H_{c}$, with
soliton peaks well separated by ''dimerized'' regions. High above $H_{c}$
(bottom panels) these regions disappear and solitons strongly overlap making
the profile more sinusoidal.}
\label{Fig2}
\end{figure}

Knowing the hyperfine and quadrupolar tensors experienced by the copper
nuclei in CuGeO$_{3}$ \cite{Fagot2}, the resonance frequency of a given
copper nucleus at site $i$ on the chain can be directly and precisely
converted into the time averaged polarization $\left\langle
S_{z}(i)\right\rangle $ of the electronic spin located on that site. To the
spatial distribution of $\left\langle S_{z}(i)\right\rangle $ in the IC
phase corresponds directly a distribution of local fields $%
A_{cc}\left\langle S_{z}(i)\right\rangle $, where $A_{cc}$ is the hyperfine
coupling constant. This gives rise to a particular NMR lineshape which is
nothing but the histogram, or the density distribution of $\left\langle
S_{z}(i)\right\rangle $ in absolute units of $\left\langle
S_{z}\right\rangle $, thus allowing a determination of the absolute value of
the order parameter. For a symmetric and periodic 1-D modulation of the spin
polarization, as expected in the IC phase of a SP system, a simple
integration of the density distribution directly provides the real space
profile over half a period\cite{Fagot1}. The NMR lineshape presented in Fig. 
\ref{Fig1} have however been analyzed somewhat differently, {\it i.e.}
fitted to the theoretically expected profile of the soliton lattice. In the
continuum approximation, the system is described by a phase variable which
satisfies a sine-Gordon equation \cite{Nakano1,Chakravarty}. The periodic
solutions are given in terms of the Jacobi elliptic functions: sn for a
lattice distortion and dn (cn) for the average (staggered) component of the
spin polarization. The total spin polarization can then be written as: 
\[
\left\langle S_{z}(i)\right\rangle =\frac{W}{2}\left\{ \frac{1}{R}\;\text{dn}%
[ic/(k\xi ),k]+(-1)^{i}\text{cn}[ic/(k\xi ),k]\right\} 
\]
where $i$ counts the position of the spins in the chain along the $c$
direction, $\xi $ is the correlation length defined by the corresponding
sine-Gordon equation, and $R$ is the ratio of amplitudes of the staggered
and the average component of spin polarization. $k$ is the modulus of
elliptic functions, and $W/2$ is the amplitude of the staggered component,
so that $W$ is just equal to the full span $\left\langle S_{z}\right\rangle
_{\max } - \left\langle S_{z}\right\rangle _{\min }$, i.e., to the total
width of the NMR spectrum. Here the frequency or magnetic field axis of the
spectra is converted to the spin polarization using the known hyperfine
coupling constant A$_{cc}$ = 4.6 T \cite{Fagot2}. When the width of the
spectra is known and their integral normalized to unity, the shape of the
spectra is determined by only two parameters $R$ and $k$. For each
experimental spectrum, these have been determined by the least square fit of
the theoretical shape of the spin density distribution (convoluted with a
convenient intrinsic NMR linewidth to allow for an experimental resolution)
as shown in Fig. \ref{Fig2}.

In addition, the period $L$ can be deduced from the NMR line using the fact
that it is directly related to the average magnetization $m=g\mu _{B}%
\overline{\left\langle S_{z}\right\rangle }$ by the relationship $\overline{%
\left\langle S_{z}\right\rangle }L=1$. Note that this is radically different
from usual NMR study of incommensurate phases in dielectrics \cite{Blinc} or
CDW systems\cite{Berthier}, in which neither the absolute value of the order
parameter nor the period can be determined. The summation of $\left\langle
S_{z}(x)\right\rangle $ over a period leads to the spatial average spin
value $\overline{\left\langle S_{z}\right\rangle }=\pi W/4K(k)R$, where $%
K(k)$ is the complete elliptic integral. In Fig. \ref{Fig3} is shown $%
\overline{\left\langle S_{z}\right\rangle }$ as a function of the magnetic
field, together with $1/L$ directly measured by X-rays, corrected for the
anisotropy of the Lande tensor $g$, since $1/L\propto 1/g$ \cite{Kiryukhin}.
The agreement is very good indeed. However, the range of magnetic field
available for X-rays was limited to approximately 4\% above $H_{c}$, and
only the vicinity of the C-IC transition could be explored in pure CuGeO$_{3}
$ samples. $\overline{\left\langle S_{z}\right\rangle }$ is found to be
proportional to $1/\ln [8/(H/Hc-1)]$, a function which is raising very
sharply at $H_{c}$ and then rapidly becoming proportional to $H$. This is
indeed predicted within the mean field theory description \cite{Buzdin} and
was already found experimentally in the magnetization measurements in other
spin-Peierls compounds \cite{Bray}.

In addition to the period $L$, the reconstitution of the spin polarization
profile as shown in Fig. \ref{Fig2} allows the determination of the
correlation length $\xi =L/[4kK(k)]$ . As a function of magnetic field $\xi $
is somewhat decreasing (Fig. \ref{Fig4}). In the field range 1.1-2 $H_{c}$,
the values of $\xi /c$ are found equal to 8-7 in agreement with the
theoretical predictions. As expected, the amplitude of the spin polarization
within the soliton,{\it \ i.e.} the width $W$ is only slightly dependent on
magnetic field (Fig. \ref{Fig4}). But the value of the ratio $R$ between
the amplitudes of the staggered and the averaged components, which is found
to decrease from 1.4 at $H_{c}$ to 1 at about 2$H_{c}$, is definitely
several times smaller than that predicted either by analytical \cite{Nakano1}
or numerical solutions \cite{Uhrig1,Meurdesoif} 
of corresponding model Hamiltonians for
an isotropic exchange. Although these results are closer to the predictions
corresponding to the XY model \cite{Fujita}, there is no reason to doubt that
the CuGeO$_{3}$ system is in the Heisenberg limit. Note also that in organic
spin-Peierls compounds, the few NMR data available in the incommensurate
phase \cite{Brom,Brown} have been interpreted as evidence for a staggered
component much larger than the average one, leading to a symmetic lineshape
broadening. However, the absence of precise knowledge of the hyperfine
coupling, and the poor definition of the lineshapes have precluded more
detailed analysis so far.

This strong reduction of the staggered component of spin polarization as
observed by NMR and compared to theoretical predictions strongly suggests,
if these latter are correct, the existence of some averaging process which
is not clearly identified yet. Since a static filtering due to a competition
between on-site and first-neighbors transferred hyperfine field can be
excluded\cite{Horvatic}, it sounds natural to consider some dynamical
process, remembering that NMR only measures time averaged values of the spin
polarization. It was recently suggested by Uhrig {\it et al}.\cite{Uhrig1}
that this averaging could be due to the zero point motion of phasons, which
are acoustic-like excitations always present in incommensurate structures.
What is the amplitude of the pinning gap for these excitations, and how they
should influence the nuclear relaxations times and the lineshapes in the IC
phase remains to be investigated.

A salient result of the analysis is the magnetic field dependence of the
shape of the soliton lattice, which is described by the modulus $k$. It
turns out that the data for the complementary modulus $k^{\prime }$ follow a
power law as a function of reduced field and up to 2$H_{c}$ (our highest
experimentally accessible value): $k^{\prime
}=(1-k^{2})^{1/2}=0.56(H/H_{c}-1)^{0.35}$, which remains to be compared to
theoretical predictions. Our experimental $k(H)$ dependence is in clear
disagreement to the old prediction based on the Hartree-Fock approximation 
\cite{Buzdin}. This is not surprising, as the customary approximation was to
take the Hartree-Fock parameter site-independent, thus reducing the full
Heisenberg Hamiltonian to the XY model, which yields to the strong
modification of the soliton lattice profile. This error has been realized
only recently during the analysis of the CuGeO$_{3}$ NMR data \cite{Foerster}%
, and true site-dependent Hartree-Fock approximation approaches indeed the
(more) exact solutions, given either by numerics or by the analytical
continuum description using bosonization\ technique.

In conclusion, we have followed the evolution of the soliton lattice as a
function of the magnetic field $H$ in the incommensurate magnetic phase of
pure CuGeO$_{3}$ by recording Cu NMR lineshapes in the field range 13-26
Tesla, i.e. for $h=(H-H_{c})/H_{c}$ varying from 0.0015 to 2. Experimental
data have been successfully fitted to the analytical solutions of the
continuum description, providing for each value of the field both the period 
$L=1/\overline{\left\langle S_{z}\right\rangle }$ of the modulation and the
width of the solitons $\xi $ \cite{remark1}. This provides a good overall
confirmation of the continuum description of quantum spin 1/2 chains. The
amplitude of the staggered magnetization is however found far smaller than
predicted by theoretical models, which could be due to the zero point motion
of phasons. The \ magnetic field dependence of $\xi $ should also be a
stringent test for numerical or analytical solutions of the spin-Peierls
Hamiltonian in presence of frustrating second neighbors coupling and 3D
interactions.

We thank S. Buzdin, D. Foerster, D. Poilblanc and G. Uhrig for stimulating
discussions. The Laboratoire de Spectrom\'{e}trie Physique is associ\'{e} au
Centre National de la Recherche Scientifique (UMR 5588) and the GHMFL is
Laboratoire conventionn\'{e} aux Universit\'{e}s J. Fourier et INPG Grenoble
I.

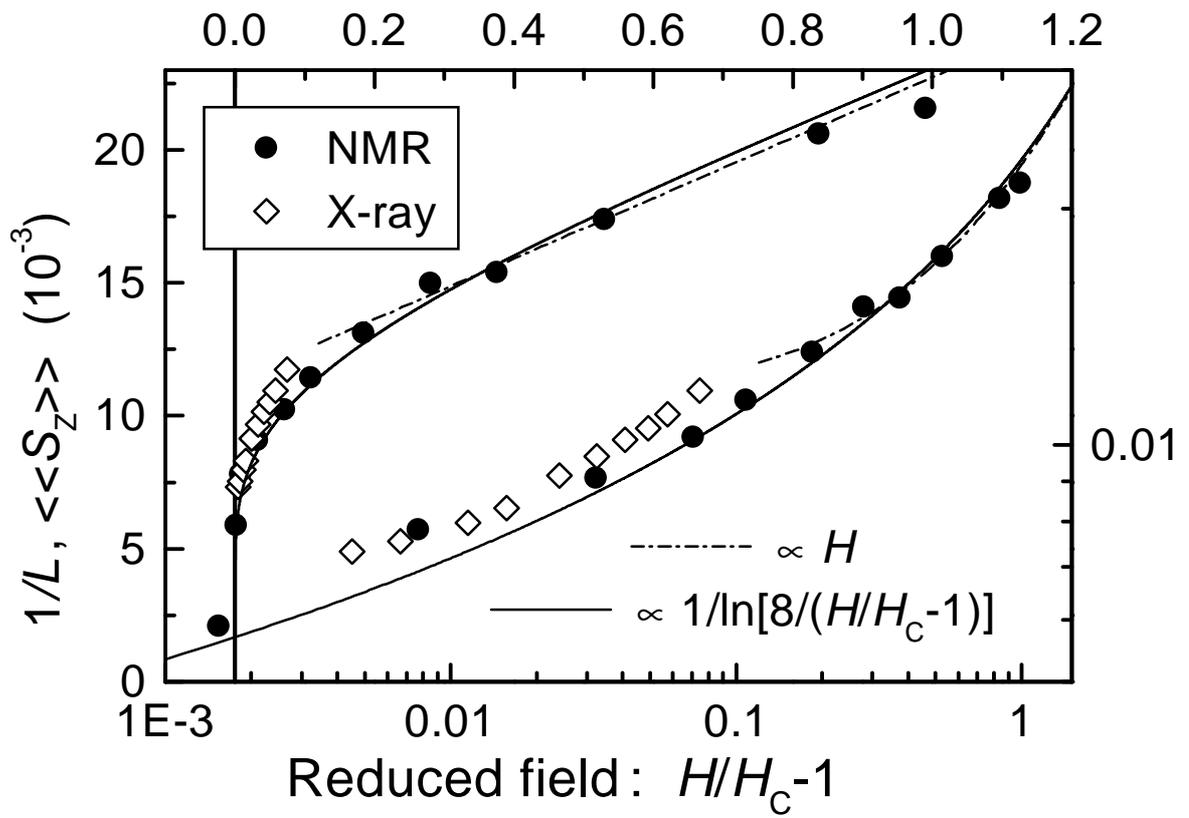
\begin{figure}[tbp]
\caption{Field dependence of the period $L$ as deduced directly from $^{65}$%
Cu NMR measurements, compared to the data available from X-rays scattering 
\protect\cite{Kiryukhin}. For convenience, the same data have been plotted
in a linear (left and top) and a logarithmic (right and bottom) scale. }
\label{Fig3}
\end{figure}
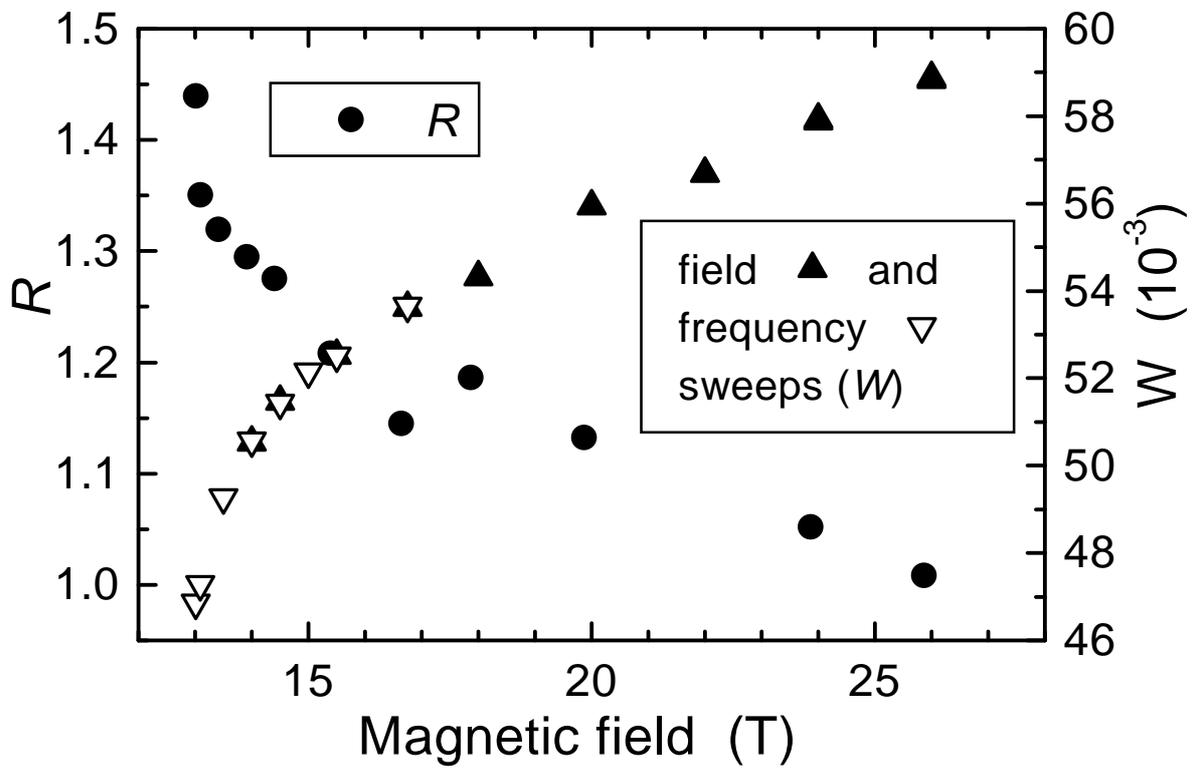
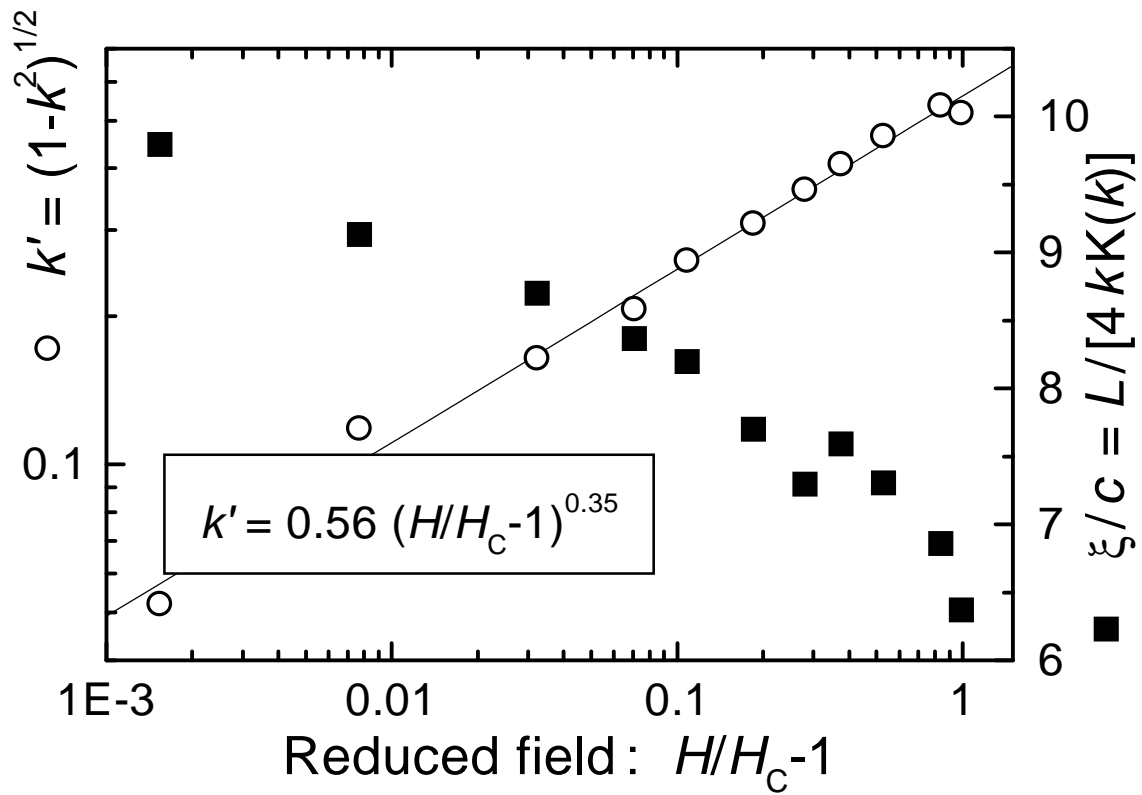
\begin{figure}[tbp]
\caption{Magnetic field dependence of the fit parameters for the NMR
lineshapes shown in Fig. \ref{Fig1}. Top panel shows the full width $W$
(right scale, triangles) of the spectra and the ratio $R$ (left scale,
circles) of amplitudes of the staggered (cn) and the average (dn) component
of the spin polarization. Bottom panel shows the correlation length $\protect%
\xi $ (right scale, squares) and the modulus $k^{\prime}$ (left scale,
circles) of Jacobi elliptic functions reflecting the evolution of the
lattice from the extreme soliton limit to a more sinusoidal profile.}
\label{Fig4}
\end{figure}


\begin{references}
\bibitem{remark1}  or alternatively the parameter $k$ of the Jacobi elliptic
functions dn or cn$(x/k\xi ,k)$, where $\xi $ is then equal to $L/4kK(k)$.
\end{references}
\end{document}